\documentclass{ws-p8-50x6-00}

\begin{document}



\thispagestyle{empty}
 
\begin{flushright}
CERN-TH/99-95\\
hep-ph/9904313
\end{flushright}
 
\vspace{1.5cm}
 
\boldmath
\begin{center}
\Large\bf A Critical Look at $\gamma$ Determinations\\ 
\vspace{0.3truecm}
from $B\to\pi K$ Decays
\end{center}
\unboldmath 

\vspace{1.2cm}
 
\begin{center}
Robert Fleischer\\
{\sl Theory Division, CERN, CH-1211 Geneva 23, Switzerland}
\end{center}
 
\vspace{1.3cm}

\begin{center}
{\bf Abstract}\\[0.3cm]
\parbox{11cm}{
The determination of the angle $\gamma$ of the unitarity triangle of 
the CKM matrix is a challenge for the $B$-factories. In this context, 
$B\to\pi K$ decays received a lot of attention, providing various 
interesting ways to constrain and determine $\gamma$. These strategies
are briefly reviewed, and their virtues and weaknesses are compared with 
one another.}
\end{center}
 
\vspace{1.5cm}
 
\begin{center}
{\sl Invited talk given at the\\
17th International Workshop on Weak Interactions and Neutrinos (WIN99),\\ 
Cape Town, South Africa, 24--30 January 1999\\ 
To appear in the Proceedings}
\end{center}
 
\vspace{1.5cm}
 
\vfil
\noindent
CERN-TH/99-95\\
April 1999
 

\vbox{}

\newpage

\thispagestyle{empty}

\mbox{}

\newpage
 
\setcounter{page}{1}
 

\title{A Critical Look at \boldmath$\gamma$\unboldmath\ Determinations\\
from \boldmath$B\to\pi K$\unboldmath\ Decays}

\author{Robert Fleischer}

\address{Theory Division, CERN, CH-1211 Geneva 23, Switzerland\\E-mail: 
Robert.Fleischer@cern.ch}


\maketitle

\abstracts{
The determination of the angle $\gamma$ of the unitarity triangle of 
the CKM matrix is a challenge for the $B$-factories. In this context, 
$B\to\pi K$ decays received a lot of attention, providing various 
interesting ways to constrain and determine $\gamma$. These strategies
are briefly reviewed, and their virtues and weaknesses are compared with 
one another.}

\section{Setting the Scene}
In order to obtain direct information on the angle $\gamma$ of the
unitarity triangle of the CKM matrix in an experimentally feasible way, 
$B\to\pi K$ decays appear very promising. Fortunately, experimental data 
on these modes are now starting to become available. In 1997, the 
CLEO collaboration reported the first results on the decays 
$B^\pm\to\pi^\pm K$ and $B_d\to\pi^\mp K^\pm$; last year, the first 
observation of $B^\pm\to\pi^0K^\pm$ was announced.\cite{stone}\, 
So far, only results for CP-averaged branching ratios have been reported, 
with values at the $10^{-5}$ level and large experimental uncertainties. 
However, already such CP-averaged branching ratios may lead to highly 
non-trivial constraints on $\gamma$.\cite{fm2}\, The following three 
combinations of $B\to\pi K$ decays were considered in the literature: 
$B^\pm\to\pi^\pm K$ and 
$B_d\to\pi^\mp K^\pm$,\cite{fm2}$^{\mbox{-}}$\cite{groro} 
$B^\pm\to\pi^\pm K$ and 
$B^\pm\to\pi^0 K^\pm$,\cite{grl}$^{\mbox{-}}$\cite{BF} 
as well as the combination of the neutral decays
$B_d\to\pi^0 K$ and $B_d\to\pi^\mp K^\pm$.\cite{BF}\, 

\section{Probing \boldmath$\gamma$ with $B^\pm\to\pi^\pm K$ and 
$B_d\to\pi^\mp K^\pm$\unboldmath}
Within the framework of the Standard Model, the most important contributions
to these decays originate from QCD penguin topologies. Making use of the 
$SU(2)$ isospin symmetry of strong interactions, we obtain
\begin{equation}\label{rel1}
A(B^+\to\pi^+K^0)\equiv P,\quad A(B_d^0\to\pi^-K^+)=-\,
\left[P+T+P_{\rm ew}^{\rm C}\right],
\end{equation}
where 
\begin{equation}
T\equiv|T|e^{i\delta_T}e^{i\gamma} \quad\mbox{and}\quad
P_{\rm ew}^{\rm C}\equiv-\,\left|P_{\rm ew}^{\rm C}\right|
e^{i\delta_{\rm ew}^{\rm C}}
\end{equation}
are due to tree-diagram-like topologies and electroweak (EW) penguins,
respectively. The label ``C'' reminds us that only ``colour-suppressed''
EW penguin topologies contribute to $P_{\rm ew}^{\rm C}$. Making use of 
the unitarity of the CKM matrix and applying the Wolfenstein parametrization 
yields
\begin{equation}
P\equiv A(B^+\to\pi^+K^0)=-\left(1-\frac{\lambda^2}{2}\right)\lambda^2A\left[
1+\rho\,e^{i\theta}e^{i\gamma}\right]{\cal P}_{tc}\,,
\end{equation}
where
\begin{equation}
\rho\,e^{i\theta}=\frac{\lambda^2R_b}{1-\lambda^2/2}
\left[1-\left(\frac{{\cal P}_{uc}+{\cal A}}{{\cal P}_{tc}}\right)\right],
\end{equation}
and $\lambda\equiv |V_{us}|$, $A\equiv|V_{cb}|/\lambda^2$, 
$R_b\equiv|V_{ub}/(\lambda V_{cb})|$. Note that $\rho$ is strongly 
CKM-suppressed by $\lambda^2R_b\approx0.02$. In the parametrization 
of the $B^\pm\to \pi^\pm K$ and $B_d\to\pi^\mp K^\pm$ observables, 
it turns out to be very useful to introduce 
\begin{equation}
r\equiv\frac{|T|}{\sqrt{\langle|P|^2\rangle}}\,,\quad\epsilon_{\rm C}\equiv
\frac{|P_{\rm ew}^{\rm C}|}{\sqrt{\langle|P|^2\rangle}}\,,
\end{equation}
with $\langle|P|^2\rangle\equiv(|P|^2+|\overline{P}|^2)/2$, as well
as the strong phase differences
\begin{equation}
\delta\equiv\delta_T-\delta_{tc}\,,\quad\Delta_{\rm C}\equiv
\delta_{\rm ew}^{\rm C}-\delta_{tc}\,.
\end{equation}
In addition to the ratio 
\begin{equation}\label{Def-R}
R\equiv\frac{\mbox{BR}(B_d\to\pi^\mp K^\pm)}{\mbox{BR}(B^\pm\to\pi^\pm K)}
\end{equation}
of CP-averaged $B\to\pi K$ branching ratios, also the ``pseudo-asymmetry'' 
\begin{equation}
A_0\equiv\frac{\mbox{BR}(B^0_d\to\pi^-K^+)-\mbox{BR}(\overline{B^0_d}\to
\pi^+K^-)}{\mbox{BR}(B^+\to\pi^+K^0)+\mbox{BR}(B^-\to\pi^-\overline{K^0})}
\end{equation}
plays an important role to probe $\gamma$. Explicit expressions for $R$ 
and $A_0$ in terms of the parameters specified above are given in Ref.\ 8. 
So far, the only available experimental result from the CLEO collaboration
is for $R$:\cite{stone}
\begin{equation}\label{RFM-exp}
R=0.9\pm0.4\pm0.2\pm0.2,
\end{equation}
and no CP-violating effects have been reported. However, if in addition 
to $R$ also the pseudo-asymmetry $A_0$ can be measured, it is possible to 
eliminate the strong phase $\delta$ in the expression for $R$, and to 
fix contours in the $\gamma\,$--$\,r$ plane,\cite{defan}\, which correspond 
to the mathematical implementation of a simple triangle 
construction.\cite{PAPIII}\, In order to determine $\gamma$, the quantity 
$r$, i.e.\ the magnitude of the ``tree'' amplitude $T$, has to be fixed. 
At this step, a certain model dependence enters. Since the properly 
defined amplitude $T$ does not receive contributions only from 
colour-allowed ``tree'' topologies, but also from penguin and annihilation 
processes,\cite{defan,bfm} it may be shifted sizeably from its 
``factorized'' value. Consequently, estimates of the uncertainty of $r$ 
using the factorization hypothesis, yielding typically 
$\Delta r={\cal O}(10\%)$, may be too optimistic.

Interestingly, it is possible to derive bounds on $\gamma$ that do {\it not}
depend on $r$ at all.\cite{fm2}\, To this end, we eliminate again $\delta$ 
in $R$ through $A_0$. If we now treat $r$ as a ``free'' variable, we find 
that $R$ takes the following minimal 
value:\cite{defan} 
\begin{equation}\label{Rmin}
R_{\rm min}=\kappa\,\sin^2\gamma\,+\,
\frac{1}{\kappa}\left(\frac{A_0}{2\,\sin\gamma}\right)^2\geq
\kappa\,\sin^2\gamma\,.
\end{equation}
Here, the quantity
\begin{equation}\label{kappa-def}
\kappa=\frac{1}{w^2}\left[\,1+2\,(\epsilon_{\rm C}\,w)\cos\Delta+
(\epsilon_{\rm C}\,w)^2\,\right],
\end{equation}
with $w=\sqrt{1+2\,\rho\,\cos\theta\cos\gamma+\rho^2}$, describes 
rescattering and EW penguin effects. An allowed range for $\gamma$ is 
related to $R_{\rm min}$, since values of $\gamma$ implying 
$R_{\rm exp}<R_{\rm min}$ are excluded. In particular, $A_0\not=0$
would allow us to exclude a certain range of $\gamma$ around $0^\circ$
or $180^\circ$, whereas a measured value of $R<1$ would exclude a
certain range around $90^\circ$, which would be of great phenomenological
importance. The first results reported by CLEO in 1997 gave $R=0.65\pm0.40$,
whereas the most recent update is that given in (\ref{RFM-exp}).

The theoretical accuracy of these constraints on $\gamma$ is limited both
by rescattering processes of the kind $B^+\to
\{\pi^0K^+,\pi^0K^{\ast+},\ldots\}$,\cite{FSI,neubert} and by 
EW penguin effects.\cite{groro,neubert}\, The rescattering
effects, which may lead to values of $\rho={\cal O}(0.1)$, can be controlled
in the contours in the $\gamma$--$r$ plane and the associated constraints 
on $\gamma$ through experimental data on $B^\pm\to K^\pm K$ decays, the 
$U$-spin counterparts of $B^\pm\to\pi^\pm K$.\cite{defan,BKK}\, Another 
important indicator for large rescattering effects is provided by
$B_d\to K^+K^-$ modes, for which there already exist stronger experimental 
constraints.\cite{groro-FSI}

An improved description of the EW penguins is possible if we use the 
general expressions for the corresponding four-quark operators, 
and perform appropriate Fierz transformations. Following these 
lines,\,\cite{defan,neubert} we arrive at 
\begin{equation}\label{EWP-expr1}
\frac{\epsilon_{\rm C}}{r}\,e^{i(\Delta_{\rm C}-\delta)}=
0.66\times \left[\frac{0.41}{R_b}\right]\times a_{\rm C}\,e^{i\omega_{\rm C}},
\end{equation}
where $a_{\rm C}\,e^{i\omega_{\rm C}}=a_2^{\rm eff}/a_1^{\rm eff}$ is the
ratio of certain generalized ``colour factors''. Experimental data on 
$B\to D^{(\ast)}\pi$ decays imply $a_2/a_1={\cal O}(0.25)$. However, 
``colour suppression'' in $B\to\pi K$ modes may in principle be different 
from that in $B\to D^{(\ast)}\pi$ decays, in particular in the presence of 
large rescattering effects.\cite{neubert}\, A first step to fix the hadronic 
parameter $a_{\rm C}\,e^{i\omega_{\rm C}}$ experimentally is provided by 
the mode $B^+\to\pi^+\pi^0$. Detailed discussions of the impact of 
rescattering and EW penguin effects on the strategies to probe $\gamma$ 
with $B^\pm\to\pi^\pm K$ and $B_d\to\pi^\mp K^\pm$ decays can be found in 
Refs.\ 7, 8  and 12.

\section{Probing \boldmath$\gamma$ with $B^\pm\to \pi^\pm K$ and 
$B^\pm\to\pi^0K^\pm$\unboldmath}
Several years ago, Gronau, Rosner and London proposed an interesting 
$SU(3)$ strategy to determine $\gamma$ with the help of 
$B^{\pm}\to\pi^{\pm} K$, $\pi^0K^{\pm}$, $\pi^0\pi^{\pm}$ decays.\cite{grl}\, 
However, as was pointed out by Deshpande and He,\cite{dh} this elegant 
approach is unfortunately spoiled by EW penguins, which play an important 
role in several non-leptonic $B$-meson decays because of the large top-quark 
mass.\cite{rf-ewp}\, Recently, this approach was resurrected by Neubert 
and Rosner,\cite{nr} who pointed out that the EW penguin contributions 
can be controlled in this case by using only the general expressions for 
the corresponding four-quark operators, appropriate Fierz transformations, 
and the $SU(3)$ flavour symmetry (see also Ref.\ 3). Since a detailed 
presentation of these strategies can be found in Ref.\ 16, we will just 
have a brief look at their most interesting features.

In the case of $B^+\to\pi^+K^0$, $\pi^0K^+$, the $SU(2)$ isospin 
symmetry implies
\begin{equation}
A(B^+\to\pi^+K^0)\,+\,\sqrt{2}\,A(B^+\to\pi^0K^+)=
-\left[(T+C)\,+\,P_{\rm ew}\right].
\end{equation}
The phase stucture of this relation, which has no $I=1/2$ piece, is
completely analogous to the $B^+\to\pi^+K^0$, $B^0_d\to\pi^-K^+$ case
(see (\ref{rel1})):
\begin{equation}
T+C=|T+C|\,e^{i\delta_{T+C}}\,e^{i\gamma},\quad
P_{\rm ew}=-\,|P_{\rm ew}|e^{i\delta_{\rm ew}}\,.
\end{equation}
In order to probe $\gamma$, it is useful to introduce observables
$R_{\rm c}$ and $A_0^{\rm c}$ corresponding to $R$ and $A_0$;\cite{BF}\,
their general expressions can be otained from those for $R$ and $A_0$ by 
making the following replacements:
\begin{equation}
r\to r_{\rm c}\equiv\frac{|T+C|}{\sqrt{\langle|P|^2\rangle}}\,, \quad
\delta\to \delta_{\rm c}\equiv\delta_{T+C}-\delta_{tc}\,,\quad
P_{\rm ew}^{\rm C}\to P_{\rm ew}.
\end{equation}
The measurement of $R_{\rm c}$ and $A_0^{\rm c}$ allows us to fix 
contours in the $\gamma$--$r_c$ plane in complete analogy to the
$B^\pm\to\pi^\pm K$, $B_d\to\pi^\mp K^\pm$ strategy. There are, however, 
important differences from the theoretical point of view. First, the 
$SU(3)$ symmetry allows us to fix $r_c\propto|T+C|$:\cite{grl}
\begin{equation}\label{SU3-rel1}
T+C\approx-\,\sqrt{2}\,\frac{V_{us}}{V_{ud}}\,
\frac{f_K}{f_{\pi}}\,A(B^+\to\pi^+\pi^0)\,,
\end{equation}
where $r_c$ thus determined is -- in contrast to $r$ -- not affected by 
rescattering effects. Second, in the strict $SU(3)$ limit, we have\cite{nr}
\begin{equation}\label{SU3-rel2}
\left|\frac{P_{\rm ew}}{T+C}\right|\,
e^{i(\delta_{\rm ew}-\delta_{T+C})}=0.66\times
\left[\frac{0.41}{R_b}\right].
\end{equation}
In contrast to (\ref{EWP-expr1}), this expression does not involve a
hadronic parameter. 

The contours in the $\gamma$--$r_c$ plane may be affected -- in analogy 
to the $B^\pm\to\pi^\pm K$, $B_d\to\pi^\mp K^\pm$ case -- by rescattering 
effects.\cite{BF}\, They can be taken into account with the help of 
additional data.\cite{defan,BKK,FSI-recent}\, The major theoretical 
advantage of the $B^+\to\pi^+K^0$, $\pi^0K^+$ strategy with respect to 
$B^\pm\to\pi^\pm K$, $B_d\to\pi^\mp K^\pm$ is that $r_c$ and 
$P_{\rm ew}/(T+C)$ can be fixed by using only $SU(3)$ arguments. 
Consequently, the theoretical accuracy is mainly limited by non-factorizable 
$SU(3)$-breaking effects. 

\vspace*{-0.1cm}

\section{Probing \boldmath$\gamma$ with $B_d\to \pi^0 K$ and 
$B_d\to\pi^\mp K^\pm$\unboldmath}
The strategies to probe $\gamma$ that are allowed by the observables of 
$B_d\to \pi^0 K$, $\pi^\mp K^\pm$ are completely analogous to the
$B^\pm\to\pi^\pm K$, $\pi^0K^\pm$ case.\cite{BF}\, However, if we require 
that the neutral kaon be observed as a $K_{\rm S}$, we have an additional 
observable at our disposal, which is provided by ``mixing-induced'' CP 
violation in $B_d\to\pi^0K_{\rm S}$ and allows us to take into account 
the rescattering effects in the extraction of $\gamma$.\cite{BF}\, To this 
end, time-dependent measurements are required. The theoretical accuracy 
of the neutral strategy is only limited by non-factorizable $SU(3)$-breaking 
corrections, which affect $|T+C|$ and $P_{\rm ew}$.

\end{document}